\documentclass[twocolumn,showpacs]{revtex4}
\usepackage{graphicx}
\usepackage{dcolumn}
\usepackage{bm}
\begin{document}

\title{The Role of Molecular Quantum Electrodynamics\\
in Linear Aggregations of Red Blood Cells}
\author{K. Bradonji\'c, J. D. Swain, A. Widom}
\affiliation{Physics Department, Northeastern University, Boston MA USA}
\author{Y. N. Srivastava}
\affiliation{Physics Department, Northeastern University, Boston MA USA}
\affiliation{Physics Department \& INFN, University of Perugia, Perugia Italy}

\begin{abstract}
Despite the fact that red blood cells carry negative charges,
under certain conditions they form cylindrical stacks, or ``rouleaux''.
It is shown here that a form of the Casimir effect, generalizing
the more well-known van der Waals forces, can provide the necessary
attractive force to balance the electrostatic repulsion. Erythrocytes
in plasma are modelled as negatively charged dielectric disks in an
ionic solution, allowing predictions to be made about the conditions
under which rouleaux will form. The results show qualitative and
quantitative agreement with observations, and suggest new experiments
and further applications to other biological systems, colloid chemistry
and nanotechnology.
\end{abstract}


\maketitle

It has been observed for many years that, under certain conditions,
erythrocytes (red blood cells) form cylindrical stacks known as
``linear aggregations'', or ``rouleaux''\cite{Chien}. One may view blood
as an ionic plasma in which red blood cells, and some other cellular
colloidal particles, are suspended. Employing direct measurements of
colloidal particle mobility, the red blood cells have been shown to
carry a net negative charge\cite{Eylar}.

\begin{figure}[bp]
\scalebox {0.6}{\includegraphics{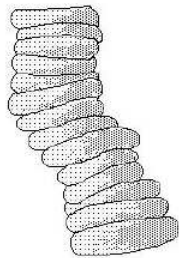}}
\caption{Although each red blood cell or slab has
a net negative charge, in certain ionic surroundings
the blood cells nevertheless tend to come in stacks
or rouleaux structures. Such a stack is schematically
drawn above.}
\label{Fig0}
\end{figure}

A long cylindrical stack of negatively charged red blood cells is
shown schematically in Fig.\ref{Fig0}. Attractive forces must exist in order
to stabilize the rouleaux formation against the explosion which would
take place if only the repulsive Coulomb interaction were
existent\cite{Chien},\cite{Eylar}. Our purpose is to argue that quantum
radiation field Casimir forces provide the required rouleaux stability
mechanism. The Casimir effect generalizes the well-known van der Waals
forces by including electromagnetic retardation effects \cite{Mostepanenko}.
Casimir forces are simpler in nature than the selective long range
dispersion forces suggested by Fr\"{o}hlich\cite{Froh, Rowlands}.

The essential physical principles underlying the Casimir forces are
quite simple. When electromagnetic radiation interacts with condensed
matter, the frequencies of the normal modes are Lamb shifted. Frequency
shifts imply a change in the electromagnetic field free energy. In particular,
at virtually zero temperature, the ground state zero point energy changes.
A special case of such an energy shift may be found from the electromagnetic
modes located between two parallel highly conducting plates. The energy shift
then describes an attractive force between two plates and can be understood
as being due to the change of zero-point modes when the distance between the
plates is changed. The closer the plates, the more reduced is the vacuum
energy and the greater the attractive force between the plates\cite{Mostepanenko}.

For the problem at hand, we consider two dielectric plates each having cross
sectional area $A$ and a dielectric constant $\varepsilon_{1}$. Between the
two dielectric plates is a plasma medium of thickness $d$ and dielectric
constant $\varepsilon_{2}$\cite{He}. The zero point electromagnetic field
energy per unit plate area is calculable. If the dielectric plates both
have a surface charge per unit area $\sigma$, then the electrostatic potential
energy density may also be computed taking into account the Debye screening
effects of a background plasma solution of ionization strength\cite{Chien} $I$.
It is assumed in both calculations\cite{VO} that $d^{2}<<A$. The final result
for the total free energy per unit area $u$ as a function of plate separation
$d$ may be written\cite{Yogi}
\begin{equation}
u(d)=\frac{\sigma^2\Lambda}{2\varepsilon_2}
\left\{e^{-d/\Lambda}-
\left(\frac{\pi^{2}\hbar \upsilon
\sqrt{\varepsilon_0\varepsilon_2}}
{360\sigma^{2}\Lambda}\right)\frac{1}{d^{3}}\right\}.
\label{free}
\end{equation}
The length $\Lambda$ describes electrolytic Debye screening and
$\upsilon$ is the electromagnetic velocity parameter
$
\upsilon=c\left[(\varepsilon_1-\varepsilon_2)
/(\varepsilon_1+\varepsilon_2)\right]^{2}
$.
The first term on the right hand side of Eq.(\ref{free}) describes the Coulomb
repulsion between red blood cells while the second term on the right
hand side of Eq.(\ref{free}) describes the Casimir attraction between red blood
cells. The relative strength of the two effects is quantitatively
described by the dimensionless parameter
\begin{equation}
a=\left(\frac{\pi^{2}\hbar \upsilon \sqrt{\varepsilon_0\varepsilon_2}}
{360\sigma^2\Lambda^{4}}\right)
=\left(\frac{\pi^{2}\hbar c \sqrt{\varepsilon_0\varepsilon_2}}
{360\sigma^2\Lambda^{4}}\right)\left[\frac{(\varepsilon_1-\varepsilon_2)}
{(\varepsilon_1+\varepsilon_2)}\right]^{2}
\label{a_parameter}
\end{equation}
where the Debye screening length \begin{math} \Lambda \end{math} is related
to the ionization strength \begin{math} I \end{math} via
\begin{math} \Lambda^2 =\{\varepsilon_2 k_BT/e^2\tilde{I}\} \end{math}.
The ionization strength in physical units is
\begin{math} \tilde{I}=\sum_a z_a^2 n_a \end{math}
where \begin{math} n_a \end{math} is the number of ions per cubic meter
having an ionic charge \begin{math} z_a |e|\end{math}. In chemical units
of moles per liter, one employs the ionization strength
\begin{math} I=[10^{-3}{\rm meter^3/liter}](\tilde{I}/N_A) \end{math}
where Avogadro's \begin{math} N_A \end{math} is the number of ions
per mole.

\begin{figure}[bp]
\scalebox {0.6}{\includegraphics{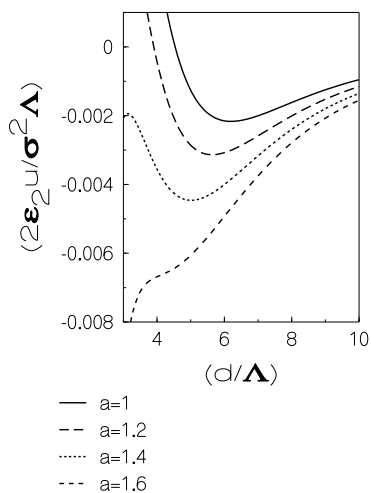}}
\caption{The variation of the free energy per unit area
with the separation distance. The rouleaux formation can exist
at the local minimum only if the parameter $a$ in
Eq.(\ref{a_parameter}) is sufficiently small.}
\label{Fig1}
\end{figure}

\begin{figure}[tp]
\scalebox {0.6}{\includegraphics{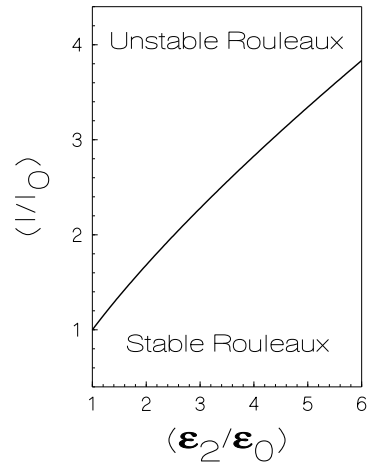}}
\caption{The phase diagram showing the regions in which
room temperature rouleaux formations are stable and in which
regions such rouleaux formations are unstable. The reference
ionization strength $I_{0}=0.05\ moles/liter$ has been employed.}
\label{Fig2}
\end{figure}

The above theory assumes that the parallel plates are maximally overlapping.
If the plates were to slide over one another yet remaining parallel but
{\em not} maximally overlapping, then the energy would be increased. Thus,
there is a force parallel to the plates tending to pull them back into a
maximally overlapping state. This force represents a {\em lateral}
Casimir effect which, though rather obvious, does not appear to have been
previously noted in the literature. It is centrally important in the context
of the rouleaux formation problem since it supplies a force which will tend
to align two plates so that they directly face each other with the largest
possible area overlap. When many plates are stacked in a cylinder,
the lateral force tends to give the cylinder transverse structural stability.
With regard to compression stability, one must find a local minimum in the
free energy per unit area $u(d)$ with respect to the distance between the
plates. A meta-stable equilibrium distance between plates arises from the
local minimum at $d_{0}$ found from equating the force per unit area to
zero; i.e. $u^\prime(d_{0})=0$. There exists a critical parameter
$a_{c}\approx{1.57}$ such that a local rouleaux minimum exists if
$a<a_{c}$ and a local rouleaux minimum does not exist if $a>a_{c}$.
The resulting free energy per unit area $u(d)$ is plotted in
Fig.\ref{Fig1} for several values of $a$.

We fix the parameters in the model so that the capacitor plates
correspond to erythrocytes and we employ room temperature. This
leaves the ionic strength $I$ and the relative to the vacuum
permeability $(\varepsilon_{2}/\varepsilon_{0})$ of the blood
plasma as free parameters. The phase diagram showing in which
regions the rouleaux are stable is then exhibited in Fig.\ref{Fig2}. To
stabilize the rouleaux it is theoretically necessary to have sufficiently
small ionic strength and/or sufficiently large relative dielectric
strength $(\varepsilon_{2}/\varepsilon_{0})$. The temperature dependence
of the Casimir energy is very weak for the problem at hand. Temperature
dependencies may be of importance in other biochemical systems.

It is important to note that rouleaux formations are usually observed
in significantly artificially diluted blood. It is also known that
such formations can be unstable in solutions of sufficiently high ionic
strength. It would be of great interest for this model to investigate
quantitatively the complete phase diagram.

In addition to providing an explanation for the otherwise mysterious
attractive force which compensates the electrostatic repulsion of
erythrocytes in rouleaux, this work opens up the possibility of numerous
experimental tests, both in blood and in novel colloids with nonspherical
particles in suspension. The fact that the Casimir energies involved are
significant suggests that studies of colloidal systems such as blood may
provide novel approaches to studies of the Casimir effect which avoid the
extremely difficult techniques employed in past
studies\cite{Sparnaay, Lamoreaux, Bressi}. It also suggests that the Casimir
effect will be of importance in the burgeoning field of nanotechnology as
devices approach the size of single cells\cite{Freitas, Drexler}.

\section*{ACKNOWLEDGMENTS}
J.D.S. would like to thank the National Science Foundation for support,
V. Voiekov for useful discussions on rouleaux formation, and
G. Hyland for discussions on Fr\"{o}hlich interactions. Y.S. would like
to thank INFN, Sezione di Perugia, Italy, for financial support. K.B.
would like to thank Eyad Katrangi for useful discussions.


\begin{thebibliography}{04}

\bibitem{Chien}
S.~Chien,in {\it The Red Blood Cell (2nd Ed.)},
edited by Douglas MacN.
Surgenor (Academic Press, New York, 1975),
Vol.2, Chap.26, p. 1032; G.V.F.~Seaman, ibid., Vol.2,
Chap. 27, p.1136.

\bibitem{Eylar}
E.H. Eylar, M.A. Madoff, O.V. Brody, and J.L. Oncley,
{\it The Journal of Biological Chemistry} {\bf 237} (6), 1992 (1962).

\bibitem{Mostepanenko}
V.M.~Mostepanenko and N.N.~Trunov,
{\it The Casimir Effect and its Applications},
translated by R.L. Znajek,(Clarendon Press, Oxford, 1997).

\bibitem{Froh}
H.~Fr\"{o}hlich, {\it Phys. Lett.} {\bf 39A}, 153 (1972).

\bibitem{Rowlands}
S.~Rowlands, L.S.~Sewchand, R.E.~Lovlin, J.S.~Beck, E.G.~Enns,
{\it Phys. Lett.} {\bf 82A}, 436 (1981).

\bibitem{He}
J.~He, A.~Karlsson, J.~Swartling,  S.~Andersson-Engels,
``{\it Light Scattering By Multiple Red Blood Cells}'',
Tech. Rep. LUTEDX/(TEAT-7117, Lund Inst. of Tech.,
Dept. of Electroscience, (2003).
{http://www.tde.lth.se/teorel/Publications/TEAT-7000-series/TEAT-7117.pdf}.

\bibitem{VO}
E.J.W.~Verwey and J.TH.G.~Overbeek,
{\it Theory of The Stability of Lyophobic Colloids},
Dover Publications, Inc., Mineola NY, (1999).

\bibitem{Yogi}
Y.~Srivastava, A.~Widom and M.~H.~Friedman,
{\it Phys. Rev. Lett.} {\bf 55}, 2246 (1985).

\bibitem{Sparnaay}
M.J.~Sparnaay, {\it Physica} {\bf{24}}, 751 (1958).

\bibitem{Lamoreaux}
S.K.~Lamoreaux, {\it Phys. Rev. Lett.} {\bf 78}, 5 (1997).

\bibitem{Bressi}
G.~Bressi, G.~Carugno, R.~Onofrio, G.~Ruoso,
{\it Phys. Rev. Lett.} {\bf{88}}, 041804-1 (2002).

\bibitem{Freitas}
Robert A. Freitas Jr. in {\it Nanomedicine},
Landes Bioscience, Austin, 1999) Vol.1.

\bibitem{Drexler}
K.~Eric Drexler in
``{\it Nanosystems: Molecular Machinery, Manufacturing, and Computation}'',
John Wiley and Sons, Inc., New York, (1992).

\end{thebibliography}
\end{document}